\documentclass[twocolumn,showpacs,aps,prl,letterpaper]{revtex4}
\usepackage{graphicx}
\usepackage{dcolumn}
\usepackage{amsmath}
\usepackage{epsfig}
\long\def\inst#1{\par\nobreak\kern 4pt\nobreak
    {\itshape #1}\par\vskip 10pt plus 3pt minus 3pt}
\RequirePackage{xspace}
\usepackage{relsize}

\begin{document}

\title{\large \bfseries \boldmath Spectrum of the charmed and b-flavored
mesons in the relativistic potential model}
\author{Jing-Bin Liu}\email{liujingbin077@mail.nankai.edu.cn}
\author{Mao-Zhi Yang}\email{yangmz@nankai.edu.cn}
\affiliation{School of Physics, Nankai University, Tianjin 300071,
P.R. China}

\date{\today}
%\date{May 31, 2013}

\begin{abstract}
We study the bound states of heavy-light quark-antiquark system in
the relativistic potential model, where the potential includes the
long-distance confinement term, the short-distance Coulomb term and
spin-dependent term. The spectrum of $B$, $B^*$, $D$, $D^*$ and states
with higher orbital quantum numbers are obtained.
Compared with previous results predicted in the relativistic potential model,
the predictions are improved and extended in this work, more theoretical masses are predicted which
can be tested in experiment in the future.
\end{abstract}

% PACS, the Physics and Astronomy Classification Scheme.

\maketitle

\section*{I Introduction}

Bound state of heavy-light quark-antiquark system $Q\bar{q}$ is of special interest. Weak decays
of such heavy-light system can be used to determine the fundamental parameters such as the
Cabibbo-Kobayashi-Maskawa (CKM) matrix elements in the standard model (SM), and to explore
the source of $CP$ violation. Experimental data from $B$ factories have confirmed the existence
of $CP$ violation in $B$ meson weak decays \cite{CPV1,CPV2}. Theoretically to treat the weak decays
of heavy-flavored mesons $B$ and $D$, effects of strong interactions have to be considered.
Strong interactions in $B$ decays can be separated into two parts, one of which can be calculated
perturbatively in QCD, while the other part is dynamically non-perturbative. The binding effect in
the quark-antiquark system is one of the main source of the non-perturbative dynamics. How to
treat the binding effect in QCD is still an open question at present. Before the non-perturbative
problem in QCD being completely solved, using phenomenological method to treat the bound state is
an effective way in practice.

The bound state of quark-antiquark can be described by the wave equation \cite{RGG,GI}
with an effective potential compatible with QCD. The potential shows a linear confining behavior at
large distance and a Coulombic behavior at short distance. Since the light quark in the heavy-light
system is relativistic, the wave equation is assumed to be with a relativistic kinematics.

The bound state effects in $B$ and $D$ mesons have been investigated with the relativistic potential
model previously in Refs. \cite{GI,RP1,RP2,RP3,RP4,RP5,ymz}. In the works of \cite{RP1,RP2,RP3,RP4,RP5,ymz}
the spin-dependent interactions are not included. For the heavy-light quark-antiquark system, the heavy
quark can be viewed as a static color source in the rest frame of the meson, and the light quark is
bound around the heavy quark by an effective potential. In the heavy quark limit the spin of the heavy quark decouples from the interaction \cite{HQa1,HQa2,HQa3,HQa4,HQa5,HQa6,HQa7,HQb}. For the realistic heavy-quark mass, the spin-dependent interactions can be treated as perturbative corrections. In Ref. \cite{GI}, the spin-dependent interactions were considered several decades ago. For most charmed and b-flavored mesons, the theoretical predictions are well consistent with experiment. Only some masses of states with orbital angular momentum $l\ge 1$ are approximately $100\; \mbox{MeV}$ higher than experimental measurements. Currently, with more experimental data available, the prediction to the spectrum of charmed and $b$-flavored mesons in the relativistic potential model needs to be improved.

In this work we will revisit the bound state effect in the heavy-flavored mesons.
The spectrum of $B$, $B^*$, $D$, $D^*$ and other heavy-light bound states with higher orbital angular
momentum and higher radial quantum number are studied. Comparing with the work of Ref. \cite{GI},
the details of the method of solving the relativistic wave equation are given, the mixing between more possible states are considered.
The spin-dependent potential is slightly modified, the predictions for the masses of charmed and $b$-flavored mesons
are more consistent with experimental measurements. We also give more predictions for the bound states with higher radial quantum number,
which can be tested in experiment in the future.

The paper is organized as follows. In section II, the relativistic wave equation for the heavy-light
quark-antiquark system and the effective Hamiltonian are given. In section III the wave equation is solved. Section IV is
for the numerical result and discussion. Section V is a brief summary.

\section*{II The wave equation for heavy-light system and the effective Hamiltonian}
The heavy flavor mesons $B$ and $D$ contain light quarks, which requires the wave equation describing
the heavy-light system include relativistic kinematics. The equation is a relativistic generalization
of Schr\"{o}dinger equation
\begin{equation}
H\psi(\vec{r})=E\psi(\vec{r}). \label{e1}
\end{equation}
The effective Hamiltonian can be written as
\begin{equation}
H=H_0+H^\prime, \label{e2}
\end{equation}
with
\begin{equation}
H_0=\sqrt{-\hbar^2 c^2\nabla^2_1+m_1^2}+\sqrt{-\hbar^2c^2\nabla^2_2+m_2^2}+V(r),
\end{equation}
where $\vec{r}=\vec{x}_2-\vec{x}_1$, and $\vec{x}_1$ and $\vec{x}_2$ are the coordinates of the heavy
and light quarks, respectively. The operators  $\nabla^2_1$ and $\nabla^2_2$ involve partial derivatives
relevant to the coordinates $\vec{x}_1$ and $\vec{x}_2$, respectively. $m_1$ is the mass of the heavy quark, and
$m_2$ the mass of the light antiquark. $V(r)$ is the effective potential of the strong interaction
between the heavy and light quarks. It can be taken as a combination of a Coulomb term and a linear
confining term, whose behavior is compatible with QCD at both short- and long-distance \cite{GI,Cornell1,Cornell2}
\begin{equation}
V(r)=-\frac{4}{3}\frac{\alpha_s(r)}{r}+b\;r+c.
\end{equation}
The first term is contributed by one-gluon-exchange diagram calculated in perturbative QCD. The Coulomb
term dominates the behavior of the potential at short-distance. The second term is the linear confining
term. The third term $c$ is a phenomenological constant, which is adjusted to give the correct ground
state energy level of the quark-antiquark system.

The running coupling constant in coordinate space $\alpha_s(r)$ can be obtained from the coupling
constant in momentum space $\alpha_s(Q^2)$ by Fourior transformation. It can be written in the following form \cite{GI}
\begin{equation}
\alpha_s(r)=\sum_i\alpha_i\frac{2}{\sqrt{\pi}}\int_0^{\gamma_i
r}e^{-x^2}dx,
\end{equation}
where $\alpha_i$ are free parameters fitted to make the behavior of the running coupling constant at
short distance be consistent with the coupling constant in momentum space predicted by QCD. The numerical
values of these parameters fitted in this work are $\alpha_1=0.15$, $\alpha_2=0.15$, $\alpha_3=0.20$, and $\gamma_1=1/2$, $\gamma_2=\sqrt{10}/2$, $\gamma_3=\sqrt{1000}/2$.

The second term $H^\prime$ in eq. (\ref{e2}) is the spin-dependent part of the Hamiltonian
\begin{equation}
H^\prime=H^{\rm hyp}+H^{\rm so},
\end{equation}
where $H^{\rm hyp}$ is the spin-spin hyperfine interaction, $H^{\rm so}$ is spin-orbit interaction.

The spin-spin hyperfine interaction used in this work is
\begin{eqnarray}
H^{\rm hyp}=&&\frac{32\pi}{9m_1 \tilde{m}_{2a}}\alpha_s(r)\delta_\sigma(r)\vec{s}_1\cdot\vec{s}_2
\nonumber\\
&&+\frac{4}{3}\frac{\alpha_s(r)}{m_1\tilde{m}_{2b}}\frac{1}{r^3}\left(\frac{3\vec{s}_1\cdot \vec{r}\vec{s}_2\cdot \vec{r}}{r^2}-
\vec{s}_1\cdot\vec{s}_2\right) \label{e7}
\end{eqnarray}
with
\begin{equation}
\delta_\sigma (r)=(\frac{\sigma}{\sqrt{\pi}})^3e^{-\sigma^2r^2},
\end{equation}
where the parameter $\sigma$ is taken as quark mass-dependent \cite{GI}
\begin{equation}
\sigma=\sqrt{\sigma_0^2\left(\frac{1}{2}+\frac{1}{2}\left(\frac{4m_1m_2}{(m_1+m_2)^2}\right)^4\right)
+s^2_0\left(\frac{2m_1m_2}{m_1+m_2}\right)^2},
\end{equation}
here $\sigma_0$ and $s_0$ are phenomenological parameters.

The spin-orbit interaction is
\begin{eqnarray}
H^{\rm so}&=&\frac{4}{3}\frac{\alpha_s(r)}{r^3}\left( \frac{1}{m_1}+\frac{1}{\tilde{m}_{2c}}\right)
    \left( \frac{\vec{s}_1\cdot\vec{L}}{m_1}+\frac{\vec{s}_2\cdot\vec{L}}{\tilde{m}_{2c}}\right)\nonumber\\
    &-&\frac{1}{2r}\frac{\partial V(r)}{\partial r} \left( \frac{\vec{s}_1\cdot\vec{L}}{m_1^2}
    +\frac{\vec{s}_2\cdot\vec{L}}{(\tilde{m}_{2d})^2}\right),\label{e10}
\end{eqnarray}
where $\vec{L}=\vec{r}\times\vec{P}$ is the relative orbital angular momentum between the quark and antiquark.

The spin-dependent interactions can be predicted by one-gluon-exchange forces in QCD \cite{RGG,GI}. The exact
form of $\propto \delta(\vec{r})$ for the spin-spin contact term $\vec{s}_1\cdot\vec{s}_2$ and $1/m_q$ for the
tensor term in $H^{\rm hyp}$ and spin-orbit interaction $H^{\rm so}$ are the predictions of one-gluon-exchange calculation
in the non-relativistic approximation. It is reasonable that there might be contributions of non-perturbative
dynamics in the bound state system and relativistic corrections for the light quark. In this work the form of
the spin-spin contact hyperfine interaction is replaced by an interaction with the behavior of exponential
suppression $e^{-\sigma^2r^2}$ as in Ref. \cite{higher-charmonium}, the mass of the light quark $m_2$'s in the denominators are
replaced by a set of new parameters $\tilde{m}_{2i}$, $i=a,\; b,\; c,\; d$, which include the relativistic corrections and the bound-state effect in the heavy meson. Originally in the potential calculated with one-gluon-exchange diagram \cite{RGG,GI}, the light quark mass $m_2$'s are in the places of the new parameters $\tilde{m}_{2i}$'s in eqs.(\ref{e7}) and (\ref{e10}), which are obtained with the approximation that both the heavy and light quarks are viewed as non-relativistic, the momenta of the quarks are dropped. However, the light quark in the heavy meson should be highly relativistic, dropping the momentum of the light quark is not a good approximation. In addition, there may also be bound-state effect in the spin-spin and spin-orbit interaction terms, which can not be treated by one-gluon-exchange diagram.  To include the relativistic effect for the light quark and the nonperturbative bound-state effect, we assume these effects can be effectively described by introducing a set of new parameters which replace the light quark mass in the denominator of the spin-dependent interaction terms. These new parameters are to be determined by fitting the experimental data on the spectrum of the charmed and b-flavored meson states. In the section of numerical treatment, one can find that this assumption does work. All the masses measured in experiment can be accommodated well.

\section*{III The solution of the wave equation}
Without the spin-dependent interaction, the solutions of the wave equation for pseudoscalar
and vector states of the quark-antiquark system shall be degenerate. The prediction to the masses of $B$ and
$B^*$, $D$ and $D^*$ will be the same. For the heavy quark and light antiquark system, the interaction decouples
to the heavy quark spin in the heavy quark limit \cite{HQa1,HQa2,HQa3,HQa4,HQa5,HQa6,HQa7,HQb}. Quark-spin
dependent interaction can be treated as perturbation. The masses of $B$ and $B^*$, $D$ and $D^*$ measured
in experiment support this treatment, the mass-differences of $B$ and $B^*$, $D$ and $D^*$ are only at the
order of a few percent \cite{PDG}.

We solve the eigen equation of $H_0$ at first, then treat the spin-dependent Hamiltonian $H^\prime$ in the
perturbation theory. The effect of $H^\prime$ will be considered to the first order in the perturbative
expansion. Denote the eigenfunction and eigenvalue of the Hamiltonian $H_0$ by $\psi^{(0)}(\vec{r})$ and
$E^{(0)}$, respectively, then the eigen equation of $H_0$ is
\begin{eqnarray}
&&\left [\sqrt{- \hbar^2c^2\nabla^2_1+m_1^2}+\sqrt{-\hbar^2c^2
\nabla^2_2+m_2^2}+V(r) \right ]\psi^{(0)}(\vec{r})\nonumber\\
&& \hspace{1cm}=E^{(0)}\psi^{(0)}(\vec{r}). \label{e11}
\end{eqnarray}
To solve the above equation, we express the wave function in terms of spectrum integration
\begin{eqnarray}
&&\psi^{(0)}(\vec{r})=\int d^3r^{\prime} \delta
^3(\vec{r}-\vec{r}^{\;\prime})\psi^{(0)}(\vec{r}^{\;\prime})\nonumber\\
&&=\int d^3r^{\prime}\int \frac{d^3k}{(2\pi \hbar c)^3}
e^{i\vec{k}\cdot(\vec{r}-\vec{r}^{\;\prime})/\hbar c}\psi^{(0)}(\vec{r}^{\;\prime}).
\label{e12}
\end{eqnarray}
With the above expression, the wave equation becomes
\begin{eqnarray}
\int\frac{d^3k}{(2\pi \hbar c)^3}d^3r^{\prime} (\sqrt{k^2
+m_1^2}+\sqrt{k^2+m_2^2} \;
)&&\nonumber\\
\times e^{i\vec{k}\cdot(\vec{r}-\vec{r}^{\;\prime})/\hbar c}
\psi^{(0)}(\vec{r}^{\;\prime})=(E^{(0)}-V(r))\psi^{(0)}(\vec{r}). &&\label{e13}
\end{eqnarray}
The exponential $e^{i\vec{k}\cdot\vec{r}/\hbar c}$ can be decomposed in terms of spherical harmonics
\begin{equation}
e^{i\vec{k}\cdot\vec{r}/\hbar c}=4\pi
\sum_{ln}i^lj_l(\frac{kr}{\hbar c})Y^*_{ln}(\hat{k})Y_{ln}(\hat{r}),\label{e14}
\end{equation}
where $j_l$ is the spherical Bessel function, $Y_{ln}(\hat{r})$ is the spherical harmonics, and $\hat{r}$ the unit
vector along the direction of $\vec{r}$.  The spherical harmonics satisfies the normalization condition
\begin{equation}
\int d\Omega
Y_{l_1n_1}(\hat{r})Y_{l_2n_2}(\hat{r})=\delta_{l_1l_2}\delta_{n_1n_2}.
\end{equation}
Using eq.(\ref{e14}) and factorize the wave function into radial and angular parts
\begin{eqnarray}
\psi^{(0)}(\vec{r})=\Phi_l(r)Y_{ln}(\hat{r}),\label{e16}
\end{eqnarray}
eq. (\ref{e13}) is transformed to be
\begin{eqnarray}
V(r)\Phi_l(r)+\frac{2}{\pi(\hbar c)^3}\int dk k^2\int
dr^{\prime}r^{\prime 2}(\sqrt{k^2 +m_1^2}&&\nonumber\\
+\sqrt{k^2+m_2^2}
)j_l(\frac{kr}{\hbar c})j_l(\frac{kr^{\prime}}{\hbar c})\Phi_l(r^{\prime})
=E^{(0)}\Phi_l(r).&& \label{e17}
\end{eqnarray}
Define a reduced radial wave function $u_l(r)$ by
\begin{equation}
\Phi_l(r)=\frac{u_l(r)}{r}, \label{e18}
\end{equation}
then the wave equation becomes
\begin{eqnarray}
V(r)u_l(r)+\frac{2}{\pi(\hbar c)^3}\int dk k^2\int
dr^{\prime}r r^{\prime}(\sqrt{k^2 +m_1^2}&&\nonumber\\
 +\sqrt{k^2+m_2^2}\;
)j_l(\frac{kr}{\hbar c})j_l(\frac{kr^{\prime}}{\hbar c})u_l(r^{\prime})
=E^{(0)}u_l(r). &&\label{e19}
\end{eqnarray}
As explained in Ref. \cite{ymz}, for a bound state of quark and antiquark, when the distance between them
is large enough, the wave function will drop seriously. Eventually the wave function will effectively vanish at a typically
large distance. We assume such a typical distance is $L$, then the quark and antiquark in bound state can be
viewed as if they are restricted in a limited space, $0<r<L$. In the limited space the reduced wave function
$u_l(r)$ for angular momentum $l$ can be expanded in terms of the spherical Bessel function
\begin{equation}
u_l(r)=\sum_{n=1}^\infty c_n \frac{a_n r}{L}j_l(\frac{a_n r}{L}), \label{e20}
\end{equation}
where $c_n$'s are the expansion coefficients, $a_n$ the $n$-th root of the spherical Bessel function $j_l(a_n)=0$.
In practice the above summation can be truncated to a large enough integer $N$
\begin{equation}
u_l(r)=\sum_{n=1}^N c_n \frac{a_n r}{L}j_l(\frac{a_n r}{L}). \label{e21}
\end{equation}
In the limited space, the momentum $k$ will be discrete. From the argument of $j_l(\frac{a_n r}{L})$ in eq. (\ref{e20}),
one can see the relevance
\begin{equation}
\frac{a_n r}{L} \Longleftrightarrow \frac{k r}{\hbar c}.
\end{equation}
Then the momentum is discretized, and the integration over $k$ in eq. (\ref{e19}) should be replaced by a summation
\begin{equation}
\frac{k}{\hbar c}\to \frac{a_n}{L}, \hspace{0.5cm}
\int \frac{dk}{\hbar c}\to \sum_n\frac{\Delta a_n }{L}, \label{e23}
\end{equation}
where $ \Delta a_n=a_n-a_{n-1}$.

Considering the limited space $0<r,\; r^\prime <L$, the discrete momentum of eq.(\ref{e23}), and substituting
eq.(\ref{e21}) into eq. (\ref{e19}), and simplify it, one can finally obtain the equation for the coefficients $c_n$'s
\begin{eqnarray}
&&\sum_{n=1}^{N}\frac{a_n}{N_m^2a_m}\int_0^L dr V(r)r^2 j_l(\frac{a_m r}{L}) j_l(\frac{a_n r}{L})c_n
+\frac{2}{\pi L^3}\Delta a_m \nonumber\\
&&\cdot a_m^2 N_m^2(\sqrt{(\frac{a_m \hbar c}{L})^2 +m_1^2}
 +\sqrt{(\frac{a_m \hbar c}{L})^2+m_2^2}\;)c_m\nonumber\\
&&
=E^{(0)}c_m, \label{e24}
\end{eqnarray}
where $N_m$ is the module of the spherical Bessel function
\begin{equation}
N_m^2=\int_0^L dr^\prime r^{\prime 2} j_l(\frac{a_m r^\prime}{L})^2.
\end{equation}
Eq. (\ref{e24}) is the eigenstate equation in the matrix form. It can be reduced to eq.(17) in Ref. \cite{ymz}
for the case $l=0$. It is not difficult to solve this equation numerically. The solution only slightly depends on
the values of $N$ and $L$ if they are large enough. We find that when $N>50$, $L>5\;{\rm fm}$, the solution of the
wave equation will be stationary.

Next we shall discuss the contribution of the spin-dependent interaction.

The spin-dependent interaction is considered perturbatively in the basis of the $| JM,sl\rangle$ sectors. $| JM,sl\rangle$ is the
eigenvector of spin-independent Hamiltonian $H_0$, where $J$ is the total angular momentum of the bound state, $M$ the magnetic quantum
number, $s$ the total spin of the quark and antiquark, $l$ the relevant orbital angular momentum between them.
The tensor part of the hyperfine interaction $H^{\rm hyp}$ in eq.(\ref{e7}) does not conserve the orbital
angular momentum, it causes mixing between the states with different orbital angular momenta $^3L_J\leftrightarrow ^3L^{\prime}_J$,
while the spin-orbit interaction $H^{\rm so}$ in eq.(\ref{e10}) does not conserve the total quark and antiquark spin, it can cause
mixing between the states with different total spin quantum numbers $^1L_J\leftrightarrow ^3L_J$. The mass matrix elements
are calculated perturbatively in the basis of $| JM,sl\rangle$. The matrix is then diagonalized to get the mixing
eigenstates. The perturbative contribution of the spin-dependent Hamiltonian $H^\prime$ to the eigenvalues of the bound
states are given below.

 (1) The eigenvalue of pseudoscalar state

The quantum number of the pseudoscalar state is $J^P=0^-$, the total spin and orbital angular momentum are $s=0$, $l=0$,
i.e., it is $^1S_0$ state. The eigenvalue of the pseudoscalar state is calculated to be
\begin{equation}
m(0^-)=E_{l=0}^{(0)}-\frac{3}{4}\langle\psi_{l=0}^{(0)}(r)| f(r)| \psi_{l=0}^{(0)}(r)\rangle .
\end{equation}

 (2) The mass matrix of the vector state, $J^P=1^-$

Both $s=1$, $l=0$ and $s=1$, $l=2$ can construct $J^P=1^-$ state. The $^3S_1$ and $^3D_1$ states can mix through the
spin-orbit interaction. The basis for the mixing is denoted to be $|\psi_1\rangle =|^3S_1\rangle$, and $|\psi_2\rangle =|^3D_1\rangle$.
The mass matrix can be written as
\begin{equation}
H=\left( \begin{array}{cc} H_{11}& H_{12}\\ H_{21} & H_{22} \end{array}\right ), \label{mix27}
\end{equation}
the results of the matrix elements are
\begin{eqnarray}
\label{H28} H_{11}&=&E_{l=0}^{(0)}+\frac{1}{4}\langle\psi_{l=0}^{(0)}(r)| f(r)| \psi_{l=0}^{(0)}(r)\rangle,\\
\label{H29} H_{12}&=&\frac{1}{\sqrt{2}}\langle\psi_{l=0}^{(0)}(r)| g(r)| \psi_{l=2}^{(0)}(r)\rangle,\\
\label{H30} H_{21}&=& H_{12}^*,\\
\label{H31} H_{22}&=& E_{l=2}^{(0)}+\langle\psi_{l=2}^{(0)}(r)|[ \frac{1}{4}f(r) -\frac{1}{2}g(r)\nonumber\\
     &&-\frac{3}{2} h_1(r)-\frac{3}{2} h_2(r)]| \psi_{l=2}^{(0)}(r)\rangle.
\end{eqnarray}
Diagonalizing the matrix $H$, one can get the eigenvalues of the two mixing states and the mixing angle. With the matrix
elements given in eqs. (\ref{H28}) $\sim$ (\ref{H31}), the above mixing matrix (\ref{mix27}) can be easily extended to
the cases with more $|^3S_1\rangle$ and $|^3D_1\rangle$ states mixing.

 (3) The eigenvalue of the scalar state, $J^P=0^+$

For the scalar state, $J^P=0^+$, the spin and orbital angular momentum are $s=1$, $l=1$. It is the $^3P_0$ state. The
eigenvalue of the scalar state is
\begin{eqnarray}
&m(0^+)&=E_{l=1}^{(0)}+\langle\psi_{l=1}^{(0)}(r)| [\frac{1}{4}f(r)-g(r)\nonumber\\
 &&- h_1(r)-h_2(r)]| \psi_{l=1}^{(0)}(r)\rangle .
\end{eqnarray}

 (4) The mass matrix of the axial-vector state, $J^P=1^+$

The $J^P=1^+$ state is mixture of $^1P_1$ and $^3P_1$ states, both states with $s=0$, $l=1$ and $s=1$, $l=1$ can construct
the $J^P=1^+$ state. The basis for the mixing is $|\psi_1\rangle =|^1P_1\rangle$, and $|\psi_2\rangle =|^3P_1\rangle$.
The matrix elements of the mass matrix are
\begin{eqnarray}
H_{11}&=&E_{l=1}^{(0)}-\frac{3}{4}\langle\psi_{l=1}^{(0)}(r)| f(r)| \psi_{l=1}^{(0)}(r)\rangle ,\label{H33}\\
H_{12}&=&\frac{1}{\sqrt{2}}\langle\psi_{l=1}^{(0)}(r)|h_2(r)- h_1(r)| \psi_{l=1}^{(0)}(r)\rangle ,\label{H34}\\
H_{21}&=& H_{12}^*,\label{H35}\\
H_{22}&=& E_{l=1}^{(0)}+\langle\psi_{l=1}^{(0)}(r)|[ \frac{1}{4}f(r)+\frac{1}{2}g(r)\nonumber\\
&& -\frac{1}{2}h_1(r)-\frac{1}{2}h_2(r)]|\psi_{l=1}^{(0)}(r)\rangle .\label{H36}
\end{eqnarray}
With eqs. (\ref{H33}) $\sim$ (\ref{H36}), the cases with more $|^1P_1\rangle$ and $|^3P_1\rangle$ mixing states
can be obtained.

 (5) The mass matrix of the tensor state, $J^P=2^+$

The $J^P=2^+$ state is mixture of $^3P_2$ and $^3F_2$ states, both states with $s=1$, $l=1$ and $s=1$, $l=3$ can
construct the $J^P=2^+$ state. The basis for the mixing is $|\psi_1\rangle =|^3P_2\rangle$, and $|\psi_2\rangle =|^3F_2\rangle$.
The matrix elements of the mass matrix are
\begin{eqnarray}
H_{11}&=&E_{l=1}^{(0)}+\langle\psi_{l=1}^{(0)}(r)|[ \frac{1}{4}f(r)-\frac{1}{10}g(r)\nonumber\\
         &&+\frac{1}{2}h_1(r)+\frac{1}{2}h_2(r)]| \psi_{l=1}^{(0)}(r)\rangle ,\label{H37}\\
H_{12}&=&\frac{3}{5}\sqrt{\frac{3}{2}}\langle\psi_{l=1}^{(0)}(r)|g(r)| \psi_{l=3}^{(0)}(r)\rangle,\label{H38}\\
H_{21}&=& H_{12}^*,\label{H39}\\
H_{22}&=& E_{l=3}^{(0)}+\langle\psi_{l=3}^{(0)}(r)| [\frac{1}{4}f(r) -\frac{2}{5}g(r)\nonumber\\
    &&-2h_1(r)-2h_2(r)]| \psi_{l=3}^{(0)}(r)\rangle .\label{H40}
\end{eqnarray}
With eqs. (\ref{H37}) $\sim$ (\ref{H40}), the cases with more $|^3P_2\rangle$ and $|^3F_2\rangle$ mixing states
can be obtained.

In the above equations, the functions $f(r)$, $g(r)$, $h_1(r)$ and $h_2(r)$ are defined as
\begin{eqnarray}
f(r)&=&\frac{32\pi}{9m_1\tilde{m}_{2a}}\alpha_s(r)\delta_\sigma(r),\\
g(r)&=&\frac{4}{3}\frac{\alpha_s(r)}{m_1\tilde{m}_{2b}}\frac{1}{r^3},\\
h_1(r)&=&\left[\frac{4}{3}\frac{\alpha_s(r)}{r^3}\left( \frac{1}{m_1}+\frac{1}{\tilde{m}_{2c}}\right)\right.\nonumber\\
         &&\left.-\frac{1}{2r}\frac{\partial V(r)}{\partial r}\frac{1}{m_1}\right]\frac{1}{m_1},\\
h_2(r)&=&\left[\frac{4}{3}\frac{\alpha_s(r)}{r^3}\left( \frac{1}{m_1}+\frac{1}{\tilde{m}_{2c}}\right)\frac{1}{\tilde{m}_{2c}}\right.\nonumber\\
         &&\left.-\frac{1}{2r}\frac{\partial V(r)}{\partial r}\frac{1}{(\tilde{m}_{2d})^2}\right].
\end{eqnarray}

\section*{IV Numerical result and discussion}
The parameters used in this work include the quark masses, the potential parameters $b$, $c$, $\tilde{m}_{2i}$, $\sigma_0$ and
$s_0$. They are selected to fit the masses of the quark-antiquark bound states. The values we obtain by fitting are
\begin{eqnarray}
&& m_b=4.99\; {\rm GeV},\;\;\; m_c=1.59\; {\rm GeV},\nonumber\\
&&m_s=0.30\;{\rm GeV},\;\;\;  m_u=m_d=0.06\; {\rm GeV},\nonumber\\
&& b=0.16\;{\rm GeV}^2,\;\;\; c=-0.28\; {\rm GeV},\nonumber\\
&&\sigma_0=1.80\; {\rm GeV},\;\;\;s_0=1.55.
\end{eqnarray}
The values of $\tilde{m}_{2a}$, $\tilde{m}_{2b}$, $\tilde{m}_{2c}$ and $\tilde{m}_{2d}$ depend on the quark-antiquark system, they can
be written as
\begin{equation}
\tilde{m}_{2i}=\epsilon_i \tilde{m}_2,\; i=a,b,c,d.
\end{equation}
We find the values of $\epsilon_i$'s and $\tilde{m}_2$ are
\begin{equation}
(\epsilon_a,\epsilon_b,\epsilon_c,\epsilon_d)=(1.00,1.30,1.30,1.32)
\end{equation}
for $(b\bar{q})$ and $(c\bar{q})$ systems,
\begin{equation}
(\epsilon_a,\epsilon_b,\epsilon_c,\epsilon_d)=(1.00,1.10,1.10,1.31)
\end{equation}
for $(b\bar{s})$ and $(c\bar{s})$ systems, and
\begin{equation}
\tilde{m}_2=\left\{\begin{array}{ll}
 0.562\;{\rm GeV}& {\rm for}\; (b\bar{q})\; {\rm system}, \\
 0.679\;{\rm GeV}& {\rm for}\; (b\bar{s})\; {\rm system},\\
 0.412\;{\rm GeV}& {\rm for}\; (c\bar{q})\; {\rm system},\\
 0.488\;{\rm GeV}& {\rm for}\; (c\bar{s})\; {\rm system},
 \end{array}\right.
\end{equation}
here $q$ is the light quark $u$ or $d$.

The solution of the wave equation does not depend on the values of $L$ and $N$ if they are taken large enough.
Numerical calculation shows that the solution is stable when $L>5\;{\rm fm}$, $N>50$. Here we take $L=10$ fm, $N=100$.

The numerical results for $(b\bar{q})$, $(b\bar{s})$, $(c\bar{q})$ and $(c\bar{s})$ bound states with the component of
radial quantum number $n=1$ dominant are given in Table \ref{t1}. Mixings between states with appropriate
quantum numbers are considered in our calculation. We find that the theoretical calculation can accommodate
the experimental data well. In addition to the masses, the mixing states relevant to each meson is also given in this table.
The vector meson states are generally mixing states of $|^3S_1\rangle$ and $|^3D_1\rangle$. The components of $|1^3S_1\rangle$
in $B^*$, $B^*_s$, $D^*$ and $D^*_s$ are overwhelmingly dominant, while the components of
$|^3D_1\rangle$ states are tiny. The masses of vector states with $|1^3D_1\rangle$ component dominant have also been predicted,
which are shown in Table \ref{t1}.

\begin{widetext}
\begin{center}
\begin{table}[h]
\caption{Theoretical spectrum of $(b\bar{q})$, $(b\bar{s})$, $(c\bar{q})$ and $(c\bar{s})$ bound states mainly with
the radial quantum number $n=1$, and the comparison with the prediction of Ref. \cite{GI} and experimental data.
The numbers in the column labeled ``GI" are theoretical masses from Ref. \cite{GI}. The experimental masses in the
last column are PDG averages \cite{PDG}.}
 \label{t1}
\begin{tabular}{|c|c|c|c|c|c|c|}\hline\hline
     & Meson & $J^P$ & Multiplet & Mass (GeV) & GI (GeV)& Exp. (MeV) \\ \hline
 &$B$ & $0^-$ & $|1^1S_0\rangle$ & 5.27 & 5.31& $5279.25\pm 0.17$ \\ \cline{2-7}
&$B^*$&$1^-$ & $\begin{array}{cc}&0.99958|1^3S_1\rangle +0.011|1^3D_1\rangle +0.021|2^3S_1\rangle +0.011|2^3D_1\rangle\\
      &-0.010|3^3S_1\rangle -0.009|3^3D_1\rangle\end{array}$
 & 5.32 & 5.32& $5325.2\pm 0.4$ \\ \cline{4-7}
&     &      & $\begin{array}{cc} &-0.012|1^3S_1\rangle+0.9963|1^3D_1\rangle+0.018|2^3S_1\rangle+0.074|2^3D_1\rangle \\
     &+0.010|3^3S_1\rangle-0.037|3^3D_1\rangle\end{array}$ & 6.05 & &  \\ \cline{2-7}
$(b\bar{q})$&     & $0^+$ & $|1^3P_0\rangle$ & 5.68 & &  \\ \cline{2-7}
&$B_1(5721)$ & $1^+$ & $-0.519|1^1P_1\rangle +0.844|1^3P_1\rangle+0.078|2^1P_1\rangle-0.111|2^3P_1\rangle$ & 5.72 & & $5723.5\pm 2.0$ \\ \cline{4-6}
&            &       & $0.851|1^1P_1\rangle +0.524|1^3P_1\rangle+0.030|2^1P_1\rangle+0.029|2^3P_1\rangle$  & 5.74 & &                \\ \cline{2-7}
&$B_2^*(5747)$ & $2^+$ & $\begin{array}{cc}&0.995|1^3P_2\rangle -0.005|1^3F_2\rangle+0.086|2^3P_2\rangle -0.004|2^3F_2\rangle\\
            &-0.047|3^3P_2\rangle -0.003|3^3F_2\rangle\end{array}$ & 5.76 &5.8 & $5743\pm 5$ \\ \cline{2-2}\cline{4-7}
&              &       & $\begin{array}{cc}&0.006|1^3P_2\rangle +0.9995|1^3F_2\rangle-0.008|2^3P_2\rangle +0.020|2^3F_2\rangle\\
            &-0.005|3^3P_2\rangle +0.020|3^3F_2\rangle\end{array}$  & 6.33 & &            \\
\hline\hline
 &$B_s$ & $0^-$ & $|1^1S_0\rangle$ & 5.35 & 5.39 & $5366.77\pm 0.24$ \\ \cline{2-7}
&$B_s^*$&$1^-$ &  $\begin{array}{cc} &0.9995|1^3S_1\rangle -0.013|1^3D_1\rangle +0.021|2^3S_1\rangle-0.012|2^3D_1\rangle\\
     &-0.011|3^3S_1\rangle+0.010|3^3D_1\rangle\end{array}$ & 5.40 & 5.45 & $5415.4^{+2.4}_{-2.1} $ \\ \cline{4-7}
&     &      &$\begin{array}{cc} &0.015|1^3S_1\rangle+0.992|1^3D_1\rangle-0.026|2^3S_1\rangle+0.113|2^3D_1\rangle \\
     &-0.009|3^3S_1\rangle-0.050|3^3D_1\rangle\end{array}$& 6.09 & &  \\ \cline{2-7}
$(b\bar{s})$&     & $0^+$ & $|1^3P_0\rangle$ & 5.72 & &  \\ \cline{2-7}
&             & $1^+$ &$-0.540|1^1P_1\rangle +0.822|1^3P_1\rangle+0.103|2^1P_1\rangle-0.151|2^3P_1\rangle$ & 5.75 & &  \\ \cline{2-2}\cline{4-7}
&$B_{s1}(5830)$ &     &$0.834|1^1P_1\rangle +0.548|1^3P_1\rangle+0.053|2^1P_1\rangle+0.039|2^3P_1\rangle$ & 5.82& & $5829.4\pm 0.7$ \\ \cline{2-7}
&$B_{s2}^*(5840)$ & $2^+$ &  $\begin{array}{cc}&0.993|1^3P_2\rangle +0.006|1^3F_2\rangle+0.107|2^3P_2\rangle +0.005|2^3F_2\rangle\\
            &-0.054|3^3P_2\rangle +0.004|3^3F_2\rangle\end{array}$ & 5.84 & 5.88 & $5839.7\pm 0.6$ \\ \cline{2-2}\cline{4-7}
&              &       &  $\begin{array}{cc}&-0.007|1^3P_2\rangle +0.998|1^3F_2\rangle+0.013|2^3P_2\rangle +0.052|2^3F_2\rangle\\
            &+0.007|3^3P_2\rangle +0.030|3^3F_2\rangle\end{array}$& 6.36&  &             \\
\hline\hline
 &$D$ & $0^-$ & $|1^1S_0\rangle$ & 1.87 & 1.88 & $1869.62\pm 0.15$ \\ \cline{2-7}
&$D^*$&$1^-$ & $\begin{array}{cc} &0.997|1^3S_1\rangle +0.033|1^3D_1\rangle -0.056|2^3S_1\rangle-0.031|2^3D_1\rangle\\
     &-0.029|3^3S_1\rangle-0.025|3^3D_1\rangle\end{array}$& 2.01 & 2.04 & $2010.28\pm 0.13$ \\ \cline{4-7}
&     &      & $\begin{array}{cc} &-0.042|1^3S_1\rangle+0.984|1^3D_1\rangle-0.079|2^3S_1\rangle-0.137|2^3D_1\rangle \\
     &+0.037|3^3S_1\rangle-0.067|3^3D_1\rangle\end{array}$& 2.75 &2.82 &  \\ \cline{2-7}
$(c\bar{q})$& $D_0^*(2400)^0$  & $0^+$ & $|1^3P_0\rangle$ & 2.30& 2.40 & $2318\pm 29$ \\ \cline{2-7}
&                & $1^+$ &$-0.096|1^1P_1\rangle +0.986|1^3P_1\rangle-0.075|2^1P_1\rangle+0.116|2^3P_1\rangle$ & 2.40& 2.44 &     \\ \cline{4-6}
& $D_1(2420)$  &       & $0.992|1^1P_1\rangle +0.106|1^3P_1\rangle+0.022|2^1P_1\rangle-0.063|2^3P_1\rangle$& 2.41 & 2.49& $2421.3\pm 0.6$ \\ \cline{2-7}
&$D_2^*(2460)$ & $2^+$ &  $\begin{array}{cc}&0.989|1^3P_2\rangle -0.013|1^3F_2\rangle-0.130|2^3P_2\rangle +0.011|2^3F_2\rangle\\
            &-0.068|3^3P_2\rangle +0.009|3^3F_2\rangle\end{array}$ & 2.45 & 2.50 & $2464.4\pm 1.9$ \\ \cline{2-2}\cline{4-7}
&              &       &  $\begin{array}{cc}&0.016|1^3P_2\rangle +0.998|1^3F_2\rangle+0.024|2^3P_2\rangle -0.045|2^3F_2\rangle\\
            &-0.019|3^3P_2\rangle -0.037|3^3F_2\rangle\end{array}$ & 3.07 & &             \\
\hline\hline
 &$D_s^\pm$ & $0^-$ & $|1^1S_0\rangle$ & 1.96 & 1.98 & $1968.49\pm 0.32$ \\ \cline{2-7}
&$D_s^{*\pm}$&$1^-$ & $\begin{array}{cc} &0.996|1^3S_1\rangle -0.041|1^3D_1\rangle -0.056|2^3S_1\rangle +0.037|2^3D_1\rangle\\
      &-0.030|3^3S_1\rangle-0.029|3^3D_1\rangle\end{array}$ & 2.10 & 2.13 & $2112.3\pm 0.5$ \\ \cline{2-2}\cline{4-7}
&     &      & $\begin{array}{cc} &0.057|1^3S_1\rangle+0.963|1^3D_1\rangle+0.157|2^3S_1\rangle-0.187|2^3D_1\rangle \\
     &-0.042|3^3S_1\rangle+0.087|3^3D_1\rangle\end{array}$& 2.77 & 2.90 &  \\ \cline{2-7}
$(c\bar{s})$& $D_{s0}^*(2317)^0$  & $0^+$ & $|1^3P_0\rangle$ & 2.31 & 2.48 & $2317.8\pm 0.6$ \\ \cline{2-7}
& $D_{s1}(2460)$ & $1^+$ & $-0.480|1^1P_1\rangle +0.850|1^3P_1\rangle -0.111|2^1P_1\rangle +0.184|2^3P_1\rangle$  & 2.42 &2.53 &  $2459.6\pm 0.6$ \\ \cline{2-2}\cline{4-7}
& $D_{s1}(2536)$ &       & $0.869|1^1P_1\rangle +0.493|1^3P_1\rangle-0.037|2^1P_1\rangle-0.031|2^3P_1\rangle$& 2.51 & 2.57 &  $2535.12\pm 0.13$ \\ \cline{2-7}
&$D_{s2}^*(2573)$ & $2^+$ &$\begin{array}{cc}&0.984|1^3P_2\rangle -0.016|1^3F_2\rangle-0.157|2^3P_2\rangle -0.012|2^3F_2\rangle\\
            &-0.077|3^3P_2\rangle -0.010|3^3F_2\rangle\end{array}$ & 2.55&2.59 & $2571.9\pm 0.8$ \\ \cline{2-2}\cline{4-7}
&              &       & $\begin{array}{cc}&0.025|1^3P_2\rangle +0.992|1^3F_2\rangle+0.057|2^3P_2\rangle +0.091|2^3F_2\rangle\\
            &-0.032|3^3P_2\rangle +0.054|3^3F_2\rangle\end{array}$ & 3.09& &             \\
\hline
\end{tabular}
\end{table}
\end{center}
\end{widetext}

The axial vector states with $J^P=1^+$ found in experiment, such as $B_1(5721)$, $B_{s1}(5830)$, $D_1(2420)$, can be explained as mixing states of $|^1P_1\rangle$ and $|^3P_1\rangle$ states. For $(b\bar{q})$ and $(c\bar{q})$ systems, we predict two almost degenerate states, respectively. The mass difference of the two mixing states in each system is very tiny. For $B_1(5721)$, we predict two states with masses $5.72 \;\mbox{GeV}$ and $5.74 \;\mbox{GeV}$. For $D_1(2420)$, the predicted masses of the two mixing states are $2.40 \;\mbox{GeV}$ and $2.41 \;\mbox{GeV}$, which are very close. For the $J^P=1^+$ state of  $(c\bar{q})$ system, the component of state with higher radial quantum number $n=2$ is not so small, the amplitude of $|2^3P_1\rangle$ can be as large as 0.116.

The $2^+$ states can be explained as mixing states of $|^3P_2\rangle$ and $|^3F_2\rangle$. The details can be found in Table \ref{t1}. Also the mixing from state of higher radial quantum number with $n=2$ can not be completely neglected, it can be as large as 10\%.

Compared with the theoretical predictions given in Ref. \cite{GI}, mixings between more quantum states are considered in this work. For bound states of $(b\bar{q})$ and $(b\bar{s})$, the masses obtained in this work are approximately consistent with the theoretical masses given in Ref. \cite{GI}, but more predicted masses are presented in this work. For $0^+$ state of $(c\bar{q})$ and  $0^+$, $1^+$ states of $(c\bar{s})$, the masses predicted in this work are about $100\;\mbox{MeV}$ smaller than the relevant predictions given in Ref. \cite{GI}, our results are more consistent with the experimental data now. The other states predicted in this work are also consistent with experiment well.

\begin{widetext}
\begin{center}
\begin{table}[h]
\caption{Theoretical spectrum of $(b\bar{q})$, $(b\bar{s})$, $(c\bar{q})$ and $(c\bar{s})$ bound states with the radial
quantum number mainly $n=2$, and the comparison with the prediction of Ref. \cite{GI}. The results in the column labeled
 ``GI" are theoretical masses from Ref. \cite{GI}.}
 \label{t2}
\begin{tabular}{|c|c|c|c|c|}\hline\hline
     & $J^P$ & Multiplet & Mass (GeV) & GI (GeV) \\ \hline
     & $0^-$ & $|2^1S_0\rangle$ & 5.81 & 5.90 \\ \cline{2-5}
 &$1^-$ & $\begin{array}{cc} &-0.020|1^3S_1\rangle-0.017|1^3D_1\rangle+0.999|2^3S_1\rangle-0.015|2^3D_1\rangle \\
     &+0.024|3^3S_1\rangle +0.011|3^3D_1\rangle\end{array}$  & 5.85 & 5.93 \\ \cline{3-5}
 &      & $\begin{array}{cc} &-0.011|1^3S_1\rangle-0.077|1^3D_1\rangle+0.014|2^3S_1\rangle+0.994|2^3D_1\rangle \\
     &-0.029|3^3S_1\rangle-0.071|3^3D_1\rangle\end{array}$ & 6.38 & \\ \cline{2-5}
$(b\bar{q})$& $0^+$ & $|2^3P_0\rangle$ & 6.04 & \\ \cline{2-5}
  & $1^+$ & $-0.077|1^1P_1\rangle +0.112|1^3P_1\rangle-0.557|2^1P_1\rangle+0.820|2^3P_1\rangle$ & 6.10 & \\ \cline{3-5}
  &       & $-0.034|1^1P_1\rangle -0.023|1^3P_1\rangle+0.826|2^1P_1\rangle+0.561|2^3P_1\rangle$ & 6.16 & \\ \cline{2-5}
  & $2^+$ & $\begin{array}{cc}&-0.077|1^3P_2\rangle +0.009|1^3F_2\rangle+0.984|2^3P_2\rangle +0.007|2^3F_2\rangle\\
            &+0.161|3^3P_2\rangle +0.006|3^3F_2\rangle\end{array}$ & 6.18 & \\ \cline{3-5}
   &       & $\begin{array}{cc}& 0.006|1^3P_2\rangle -0.021|1^3F_2\rangle-0.011|2^3P_2\rangle +0.998|2^3F_2\rangle\\
            &+0.028|3^3P_2\rangle +0.042|3^3F_2\rangle\end{array}$  & 6.61 &  \\
\hline\hline
  & $0^-$ & $|2^1S_0\rangle$ & 5.89 & 5.98 \\ \cline{2-5}
 &$1^-$ &  $\begin{array}{cc} &-0.020|1^3S_1\rangle+0.024|1^3D_1\rangle+0.999|2^3S_1\rangle+0.017|2^3D_1\rangle \\
     &+0.025|3^3S_1\rangle-0.012|3^3D_1\rangle\end{array}$  & 5.94 & 6.01  \\ \cline{3-5}
 &      &$\begin{array}{cc} &0.013|1^3S_1\rangle-0.117|1^3D_1\rangle-0.016|2^3S_1\rangle+0.987|2^3D_1\rangle \\
     &+0.044|3^3S_1\rangle-0.097|3^3D_1\rangle\end{array}$& 6.41 & \\ \cline{2-5}
$(b\bar{s})$ & $0^+$ & $|2^3P_0\rangle$ & 6.08 & \\ \cline{2-5}
 & $1^+$ &$-0.103|1^1P_1\rangle +0.151|1^3P_1\rangle-0.550|2^1P_1\rangle+0.815|2^3P_1\rangle$ & 6.15 & \\ \cline{3-5}
  &     &$-0.055|1^1P_1\rangle -0.037|1^3P_1\rangle+0.827|2^1P_1\rangle+0.558|2^3P_1\rangle$ & 6.24 & \\ \cline{2-5}
 & $2^+$ &  $\begin{array}{cc}&-0.095|1^3P_2\rangle -0.014|1^3F_2\rangle+0.977|2^3P_2\rangle -0.008|2^3F_2\rangle\\
            &+0.192|3^3P_2\rangle -0.007|3^3F_2\rangle\end{array}$ & 6.26 & \\ \cline{3-5}
  &       &  $\begin{array}{cc}&-0.016|1^3P_2\rangle -0.053|1^3F_2\rangle+0.036|2^3P_2\rangle +0.984|2^3F_2\rangle\\
            &-0.148|3^3P_2\rangle +0.079|3^3F_2\rangle\end{array}$& 6.64 &   \\
\hline\hline
  & $0^-$ & $|2^1S_0\rangle$ & 2.46 & 2.58 \\ \cline{2-5}
 &$1^-$ & $\begin{array}{cc} &0.049|1^3S_1\rangle+0.075|1^3D_1\rangle+0.991|2^3S_1\rangle-0.050|2^3D_1\rangle \\
     &-0.072|3^3S_1\rangle-0.034|3^3D_1\rangle\end{array}$& 2.59 & 2.64  \\ \cline{3-5}
 &      & $\begin{array}{cc} &0.036|1^3S_1\rangle+0.142|1^3D_1\rangle+0.052|2^3S_1\rangle+0.970|2^3D_1\rangle \\
     &+0.153|3^3S_1\rangle+0.111|3^3D_1\rangle\end{array}$& 3.11 & \\ \cline{2-5}
$(c\bar{q})$  & $0^+$ & $|2^3P_0\rangle$ & 2.67 &  \\ \cline{2-5}
  & $1^+$ &$0.080|1^1P_1\rangle -0.130|1^3P_1\rangle-0.474|2^1P_1\rangle+0.867|2^3P_1\rangle$ & 2.83 &  \\ \cline{3-5}
 &       & $0.010|1^1P_1\rangle +0.011|1^3P_1\rangle+0.877|2^1P_1\rangle+0.480|2^3P_1\rangle$& 2.87 & \\ \cline{2-5}
 & $2^+$ &  $\begin{array}{cc}& 0.108|1^3P_2\rangle -0.028|1^3F_2\rangle+0.962|2^3P_2\rangle +0.019|2^3F_2\rangle\\
            &-0.248|3^3P_2\rangle +0.015|3^3F_2\rangle\end{array}$ & 2.93 & \\ \cline{3-5}
  &       &  $\begin{array}{cc}&-0.047|1^3P_2\rangle +0.040|1^3F_2\rangle-0.102|2^3P_2\rangle +0.929|2^3F_2\rangle\\
            &-0.344|3^3P_2\rangle +0.062|3^3F_2\rangle\end{array}$ & 3.39 &  \\
\hline\hline
   & $0^-$ & $|2^1S_0\rangle$ & 2.55& 2.67 \\ \cline{2-5}
 &$1^-$ & $\begin{array}{cc} & 0.043|1^3S_1\rangle-0.147|1^3D_1\rangle+0.982|2^3S_1\rangle+0.074|2^3D_1\rangle \\
     &-0.070|3^3S_1\rangle-0.046|3^3D_1\rangle\end{array}$ & 2.68 & 2.73 \\ \cline{3-5}
  &      & $\begin{array}{cc} &-0.045|1^3S_1\rangle+0.183|1^3D_1\rangle-0.071|2^3S_1\rangle+0.893|2^3D_1\rangle \\
     &-0.383|3^3S_1\rangle-0.125|3^3D_1\rangle\end{array}$& 3.14 & \\ \cline{2-5}
$(c\bar{s})$  & $0^+$ & $|2^3P_0\rangle$ & 2.69& \\ \cline{2-5}
 & $1^+$ & $0.112|1^1P_1\rangle -0.184|1^3P_1\rangle-0.504|2^1P_1\rangle+0.837|2^3P_1\rangle$ & 2.87 &  \\ \cline{3-5}
 &       & $0.042|1^1P_1\rangle +0.025|1^3P_1\rangle +0.856|2^1P_1\rangle +0.515|2^3P_1\rangle$ & 2.97 & \\ \cline{2-5}
  & $2^+$ &$\begin{array}{cc}& 0.127|1^3P_2\rangle -0.063|1^3F_2\rangle+0.946|2^3P_2\rangle -0.028|2^3F_2\rangle\\
            &-0.289|3^3P_2\rangle -0.020|3^3F_2\rangle\end{array}$ &3.02 &  \\ \cline{3-5}
  &       & $\begin{array}{cc}&0.000|1^3P_2\rangle -0.100|1^3F_2\rangle-0.013|2^3P_2\rangle +0.978|2^3F_2\rangle\\
            &-0.123|3^3P_2\rangle +0.136|3^3F_2\rangle\end{array}$ & 3.41 & \\
\hline
\end{tabular}
\end{table}
\end{center}
\end{widetext}

The radial excited states with the quantum number up to $n=2$ are also predicted, they are given in Table \ref{t2}. In general,
our results for $n=2$ are slightly smaller than the theoretical masses given in Ref. \cite{GI}. In addiction, more predicted masses are given in this work, which can be tested in experiment in the future.

Both Belle and BaBar collaborations found a $D_{sJ}$ resonance in the analysis of $DK$ mass distribution, $D_{sJ}(2700)$ denoted by Belle collaboration \cite{Belle0}, and $X(2690)$ denoted by BaBar collaboration \cite{BaBar}. The mass and width are
\begin{eqnarray}
M&=& 2708\pm 9^{+11}_{-10} \mbox{MeV}, \nonumber\\
\Gamma &=& 108\pm 23 ^{+36}_{-31}  \mbox{MeV}
\end{eqnarray}
measured by Belle collaboration \cite{Belle0}, and
\begin{eqnarray}
M&=& 2688\pm 4\pm 3 \mbox{MeV}, \nonumber\\
\Gamma &=& 112\pm 7\pm 36  \mbox{MeV}
\end{eqnarray}
measured by Babar collaboration \cite{BaBar}.

It is possible that $D_{sJ}(2700)$ and  $X(2690)$ are the same resonance. Comparing the masses $D_{sJ}(2700)$ and  $X(2690)$ measured by Belle and BaBar collaborations with the predicted mass for $J^P=1^-$ $(c\bar{s})$ state given in Table \ref{t2}, one can find that the state with predicted mass $2.68$ GeV is consisted with the $D_{sJ}$ meson found in experiment. Therefore $D_{sJ}(2700)$ and/or  $X(2690)$ can be identified as the first radial excitation of $D_s^*(2112)$, which agrees with Ref. \cite{CFNR} analyzed due to heavy quark limit.

It is interesting to discuss the properties of the heavy-light quark-antiquark bound states from the point view of heavy quark symmetry. The spectroscopy of mesons with open charm and beauty flavors was analyzed in the heavy quark limit recently in Ref. \cite{CFNR,CFN,CFGN}, where the mesons are classified in heavy quark doublets and the quantum numbers are assigned to the heavy flavored mesons. Here we would like to discuss how the properties of the heavy flavored mesons implied by the heavy quark limit are reproduced in the calculation in the potential model. In the heavy quark limit, the spin of the heavy quark decouples from the light degrees of freedom. The heavy quark spin $s_Q$ and the total angular momentum of the light antiquark $s_l$ conserve separately in the bound state of the heavy-light system. Therefore heavy-flavored mesons can be classified according to the values of the angular momentum of the light antiquark $s_l$. The total spins of the mesons are $J=s_l\pm \frac{1}{2}$, according to which the mesons can be collected into doublets. For any value of the orbital angular momentum of the light antiquark $l$, the parity of the meson is $P=(-1)^{l+1}$, and the total angular momentum of the light antiquark $\vec{s}_l=\vec{s}_q +\vec{l}$, where $s_q$ is the spin of the light antiquark. Since the properties of the hadronic states do not depend on the spin and flavor of the heavy quark due to the heavy quark symmetry, the mesons within the same doublet degenerate in the heavy quark limit. For $l=0$, the total angular momentum of the light antiquark is $s_l=\frac{1}{2}$, then the total spin of the $Q\bar{q}$ meson could be 0 and 1. These two states form a doublet with $J^P_{s_l}=(0^-,1^-)_{1/2}$. For $l=1$, the possible angular momenta of light antiquark are $s_l=\frac{1}{2}$ and  $s_l=\frac{3}{2}$. There are two doublets in this case, $J^P_{s_l}=(0^+,1^+)_{1/2}$ and $J^P_{s_l}=(1^+, 2^+)_{3/2}$. For each meson state with specified quantum numbers classified in the heavy quark limit, we can denote them by $|J^P\rangle ^l_{s_l}$. By analyzing angular momentum addition with the help of Clebsch-Gordan coefficients, we can decompose the states in the heavy quark limit into combination of states with definite quantum numbers $l$, $S$ and $J$, i.e., the state $|^{2S+1}L_J\rangle$. For states in the doublet $J^P_{s_l}=(0^-,1^-)_{1/2}$, we get
\begin{eqnarray}
|0^-\rangle ^{l=0}_{s_l=1/2}&=&|^1S_0\rangle ,\label{e50}\\
|1^-\rangle ^{l=0}_{s_l=1/2}&=&|^3S_1\rangle .\label{e51}
\end{eqnarray}
For states in the doublet $J^P_{s_l}=(0^+,1^+)_{1/2}$, we can obtain
\begin{eqnarray}
|0^+\rangle ^{l=1}_{s_l=1/2}&=&|^3P_0\rangle ,\label{e52}\\
|1^+\rangle ^{l=1}_{s_l=1/2}&=&-\sqrt{\frac{1}{3}}|^1P_1\rangle +\sqrt{\frac{2}{3}}|^3P_1\rangle .\label{e53}
\end{eqnarray}
For states in the doublet $J^P_{s_l}=(1^+, 2^+)_{3/2}$, the results are
\begin{eqnarray}
|1^+\rangle ^{l=1}_{s_l=3/2}&=&\sqrt{\frac{2}{3}}|^1P_1\rangle +\sqrt{\frac{1}{3}}|^3P_1\rangle ,\label{e54}\\
|2^+\rangle ^{l=1}_{s_l=3/2}&=&|^3P_2\rangle .\label{e55}
\end{eqnarray}
For the $1^-$ state in the doublet $J^P_{s_l}=(1^-, 2^-)_{3/2}$, the result is
\begin{equation}
|1^-\rangle ^{l=2}_{s_l=3/2}=|^3D_1\rangle. \label{e56}
\end{equation}
The result of $2^+$ state in the doublet $J^P_{s_l}=(2^+, 3^+)_{5/2}$ is
\begin{equation}
|2^+\rangle ^{l=3}_{s_l=5/2}=|^3F_2\rangle. \label{e57}
\end{equation}
The above eqs. (\ref{e50})-(\ref{e57}) give the results that the meson states with definite $J^P$ quantum number expanded as states of $|^{2S+1}L_J\rangle$ in the heavy quark limit. Comparing these results with the column ``Multiplet" in Table \ref{t1} and \ref{t2} for each meson, one can see that the results are consistent with the heavy quark limit, there are only small deviation from the heavy quark limit for most mesonic states. The small deviation is due to the masses of the heavy quarks $b$ and $c$ used here are realistic values, not infinity. Only the component for the $J^P=1^+$ state of non-strange $c\bar{q}$ meson is greatly different from the case in the heavy quark limit. The mixing of $|1^1P_1\rangle$ and $|1^3P_1\rangle$ in $1^+$ state of $c\bar{q}$ is very small, the mixing angle is only about $-0.10$ rad, while the mixing angle in the heavy quark limit should be $-\mbox{ArcSin}\sqrt{1/3}=-0.615\; \mbox{rad}$. We checked the reason and find that without considering the contribution of states with $n=2$, the mixing angle is indeed very close to the heavy quark limit. In addition, the eigenvalues of the two mixing states is very near, they almost degenerate, and the gap between the energy levels with radial numbers $n=1$ and $n=2$ is not large, the mixing effect of $n=2$ state will not be negligible. After adding the mixing effect of $n=2$ state, the mixing angle between $|1^1P_1\rangle$ and $|1^3P_1\rangle$ is seriously affected, which makes it very small, i.e., the mixing angle is $\theta =-0.10\; \mbox{rad}$ . The result with small mixing angle is consistent with experiment. The Belle collaboration determined the possible mixing angle between the two $1^+$ non-strange charmed meson states. They obtain a small mixing angle: $\theta =-0.10\pm 0.03\pm 0.02\pm 0.02\; \mbox{rad}$ \cite{Belle}. The theoretical prediction for the mixing angle in this work is in good agreement with experiment.

Finally the wave function of each bound state can be obtained simultaneously when solving the wave equation, which is not given
here explicitly. But it is easy to get the wave function when it is needed.
\section*{V Summary}
The bound states of heavy-light quark and antiquark system are studied in the relativistic potential model. The dynamics
of the light quark in the system requires the wave equation describing the bound state include relativistic kinematics.
The potential is compatible with QCD, it shows the behavior of Coulomb potential at short distance, and a linear confining
behavior at large distance. The spin-dependent interactions are also considered. The spectrums of $B$ and $D$ system are
obtained. Compared with the results obtained in the relativistic potential model previously, the predictions to the spectrum
are improved. The masses of the bound states with the radial quantum number $n=1$ are well consistent with the experimental
measurement. In addition, the masses of more meson states are predicted, which can be tested in experiment in the future. The
wave function of each bound state can be also obtained by solving the wave equation.

\vspace{0.5cm}

After this work is finished, we find the experimental data newly presented by LHCb collaboration \cite{LHCb}, where several
new resonances are observed in the mass region between $2500$ and $3000\; \mbox{MeV}$. Comparing with the experimental data, we find that  the resonance $D_J(2580)$ can be assigned as dominantly $|2^3S_1\rangle$ state of $(c\bar{q})$ with $J^P=1^-$, $D_J(3000)$ assigned as $|2^3P_2\rangle$ state with $J^P=2^+$, which can be seen by comparing the theoretical prediction in Table \ref{t2} with experimental data in \cite{LHCb}.

%%%%%%%%%%%%%%%%%%%%%%%%%%%%%%%%%%%%%%%%%%%%%%%%%%%%%%%%%%%%%%%%%%%%%%%%%
% ACKNOWLEDGMENTS
%%%%%%%%%%%%%%%%%%%%%%%%%%%%%%%%%%%%%%%%%%%%%%%%%%%%%%%%%%%%%%%%%%%%%%%%%

\section*{Acknowledgments} This work is supported in part by the
National Natural Science Foundation of China under contracts Nos.
11375088, 10975077, 10735080, 11125525 and by the Fundamental Research Funds for the
Central Universities No. 65030021.

%%%%%%%%%%%%%%%%%%%%%%%%%%%%%%%%%%%%%%%%%%%%%%%%%%%%%%%%%%%%%%%%%%%%%%%%%
% BIBLIOGRAPHY
%%%%%%%%%%%%%%%%%%%%%%%%%%%%%%%%%%%%%%%%%%%%%%%%%%%%%%%%%%%%%%%%%%%%%%%%

\end{document}